\documentstyle [12pt,epsf]{article}

\input{epsf}
\begin{document}
\begin{titlepage}
\begin{center}

 \vspace{-0.5in}

{\large \bf Entangled States\\ and\\
Super-radiant Phase Transition}\\
 \vspace{.2in}{\large\em M. Aparicio Alcalde\,\footnotemark[1],
 A. H. Cardenas\,\footnotemark[2], N. F. Svaiter
\footnotemark[3]}\\
\vspace{.1in}
 Centro Brasileiro de Pesquisas F\'{\i}sicas,\\
 Rua Dr. Xavier Sigaud 150,\\
 22290-180, Rio de Janeiro, RJ, Brazil. \\

\vspace{.2in}{\large\em  V. B. Bezerra\,\footnotemark[4]}\\

\vspace{.1in} Departamento de F\'{\i}sica,\\ Universidade Federal
da
Para\'{\i}ba,\\
58059-970, Jo\~ao Pessoa, PB, Brazil.\\

\subsection*{\\Abstract}
\end{center}

\baselineskip .18in

The Dicke spin-boson model is composed by a single bosonic mode
and an ensemble of $N$ identical two-level atoms. Assuming thermal
equilibrium with a reservoir at temperature $\beta^{-1}$, we
consider the situation where the coupling between the bosonic mode
and the atoms generates resonant and non-resonant processes. The
thermodynamic of the model is investigated. Next we introduce
dipole-dipole interaction between the atoms. We investigate the
transition from fluorescent to super-radiant phase and the quantum
phase transition in a situation where the dipole-dipole
interaction between the atoms generates entangled states in the
atomic system. We proved that, the critical behavior is not
modified by the introduction of the dipole-dipole interaction.

\footnotetext[1]{e-mail:\,\,aparicio@cbpf.br}
\footnotetext[2]{e-mail:\,\,cardenas@cbpf.br}
\footnotetext[3]{e-mail:\,\,nfuxsvai@cbpf.br}
\footnotetext[4]{e-mail:\,\,valdir@fisica.ufpb.br} PACS numbers:

42.50.Nn, 05.30.Jp, 73.43.Nq

\end{titlepage}
\newpage\baselineskip .18in
\section{Introduction}
\quad $\,\,$ According to classical physics, the knowledge of the
state of a composed system yields complete knowledge of the
individual state of the parts. This is the principle of
separability \cite{separability} \cite{separability2}. For
example, if we known the state of a classical system of particles,
we known the state of each particle. However, this is not true in
the context of quantum mechanics. One of the consequences of the
superposition principle is the introduction of the concept of
entanglement. After interacting, two quantum systems can end up in
a non-separable state, i.e., states that can not be factorized
into a product of the states of its sub-systems. These states of
such composite system are called entangled states
\cite{schrodinger}. For instance, in a bipartite entangled state,
each part losses its quantum identity. Entanglement leads to the
non-locality properties of quantum mechanics, and action at a
distance at a speed greater than the speed of light \cite{light0}
\cite{light}. Discussing the hidden variable theories based on
local realism, Bell has shown that quantum theory is supposed to
be non-local \cite{bell1} \cite{bell2} \cite{bell}. It was be
shown that any pure entangled state of two spin-$\frac{1}{2}$
violates corresponding Bell's inequality \cite{exp1} \cite{exp2}
\cite{exp3} \cite{mann}.

With the development of quantum information \cite{zurek1}
\cite{khalili} \cite{namiki} and its application in computation
and communication \cite{benioff} \cite{fey1} \cite{albert}
\cite{deutsch} \cite{fey2} \cite{di} \cite{steane} \cite{livro},
the entangled states have been attracted enormous interest. For
instance, new tests of quantum mechanics can be implemented using
some entangled states, as for example the Bell and the GHZ states
\cite{176} \cite{zurek}. Also, several quantum protocols can be
realized exclusively with the help of entangled states \cite{hep}
\cite{rbrandes}, as for example the quantum dense coding protocol
proposed by Bennett and Wiesner \cite{wiesner}, or the quantum
teleportation protocol proposed by Bennett and co-workers
\cite{wootters}. In the teleportation of entanglement a single
qubit in an arbitrary state can be transferred, exchanging
information over long distances, without physically transferring
the system itself. The maximally entangled pair of qubits is
crucial in the quantum teleportation. Actually, there are many
physical systems that can be used to implement a quantum computer
with the quantum logic gates. Some experimental devices are based
on cavity quantum electrodynamics, trapped ions and nuclear
magnetic resonances. For example, for the realization of quantum
logic gates, a many-body system prepared in an entangled state has
been proposed \cite{cirac}. Another promising system used to
implement quantum computation is the quantum dot array proposed by
Loss and DiVincenzo \cite{loss}. The basic problem that arises in
this area of research is how to create entangled in many-body
systems \cite{cm1} \cite{cm2} \cite{cm3} and also how to generate
systems which are not affected by the environment, overcoming the
problem of decoherence.

In the simplest case of two atoms, it is quite important to
demonstrate creation of entanglement on such system. There are in
the literature different methods for detection of entangled states
of two interacting atoms. One method is based in measure the
angular intensity distribution of the fluorescent field emitted by
these two atoms, since it is well known that radiation emitted by
atoms exhibits directional properties \cite{lehmberg1}
\cite{lehmberg2} \cite{agarval} \cite{tanas} \cite{ficek}
\cite{ujihara}. The other is based in the dynamic of the
population inversion of the system. For instance, the properties
of spontaneous emission from two identical entangled atoms
interacting with the modes of a bosonic field was investigated by
Guo and Yang \cite{guo}. They have shown that the time evolution
of the population inversion, which is proportional to the
radiation intensity, depends on the degree of entanglement of the
initial state of the system.

Suppose that in a many-body system an entanglement has been
created in a given portion of such system. A fundamental question
is how to known whether a state of a many-body system is entangled
or not \cite{amico}. Before continue, we would like to briefly
describe the theoretical machinery that are needed to define
entanglement in a pure or mixed state. To define separability and
entanglement in bipartite systems, we use the properties of the
state vector or density operator. Since a pure state is entangled
if it is not separable, it can be shown that a pure state is
separable if and only if the reduced density operators of
sub-systems represent pure states. Therefore a pure state is
entangled if and only if the reduced density operators for the
sub-systems describe mixed states. In practice, a pure state is
separable if and only if the quantum fluctuations of all
observable (linear self-adjoint operators acting in the respective
Hilbert space of each sub-system) are uncorrelated. If at least
one pair of linear self-adjoint operators have correlated quantum
fluctuations, the pure state is entangled.  For mixed states we
define a separable mixed state if the two sub-systems have the
same purity and von Neumann entropy. An entangled mixed state is
one that is not separable. We would like to stress that
experimental tests for separability and entanglement of pure
states using correlated quantum fluctuations are not available for
mixed states. Strong correlation between two operators does not
means that there is entanglement for a system described by a mixed
state.

Critical phenomena is a cooperative effect characterized by
fluctuations in the order parameter, where all scales of length
are important. For instance, the two-point correlation function
(the expectation value of products of local observables) near the
phase transition does not decrease exponentially, but
polynomially, as the distance between the points goes to infinity.
The system exhibit long range order. Since, for pure states the
correlated quantum fluctuations of two observables defines an
entangled state, we should expect a close connection between
second order phase transition at zero temperature and
entanglement. We would like to note that recently Emary and
Brandes \cite{eb} \cite{eb2} discussed the connection between
quantum phase transition and the chaotic behavior that emerges in
the full Dicke model for finite $N$, where the energy
level-spacing statistics changes from Poissonian to one described
by Gaussian ensembles of the random matrix theory \cite{brody}
\cite{mehta}.

Let us now focus our attention for the super-radiant phase
transition. Consider an ensemble of $N$ two-level atoms, all
prepared into the excited state. Each atom can emit a photon by
spontaneous emission. In the situation where there is not coupling
between the atoms, each atom radiates independently. This is
called the fluorescent phase. Let us consider the following
experimental environment: the $N$ two-level atoms are in an
optical cavity where all atoms are in resonance with a single mode
of the field. For an ensemble of atoms in a volume with linear
dimensions small compared to the emission wavelength, they start
to radiate spontaneously faster and strongly than the ordinary
fluorescent phase. In this situation the radiation rates becomes
quadratic dependent on the number of atoms. This cooperative
process is called super-radiance. Recently Lambert and co-workers
\cite{lambert2} investigated the entanglement properties of an
ensemble of $N$ two-level atoms interacting with a  field mode in
the super-radiant phase \cite{gross} via the von Neumann entropy.
The maximum entanglement occurs near the critical region. The
authors also studied the system at zero temperature and how does
the entanglement is affected by a quantum phase transition that
occurs in the system \cite{hertz} \cite{sachdev}. By the other
hand, in a system of $N$ atoms, the presence of the dipole-dipole
interaction can generates entangled states in the many-body
system. See for example Ref. \cite{brennen}, where was discussed
in detail how to generate multi-particles entanglement of atoms
trapped by harmonic potential, interacting with a classical field
and between them via the dipole-dipole interaction. Therefore one
can imagine that if entangled has been created in a portion of the
many-body system, the critical behavior of the system changes due
to the strong correlation between the atoms.

The aim of this work is to investigate if the entanglement between
two-level atoms, generated by the dipole-dipole interaction, is
able to change the critical properties of the system. For
instance, it is well known that the dipole-dipole decohering
effect is able to suppress the super-radiant emission
\cite{hartmann} \cite {coffey}. Note that we are not able to
quantify the degree of entanglement present in the model since we
are using functional integration methods to investigate the
thermodynamic of the model. First, we investigate the full Dicke
model where the quantum phase transition and a phase transition
from fluorescence to super-radiant phase, at some temperature
$\beta^{-1}$, in the system of $N$ atoms interacting with a
bosonic field is analyzed. Second, introducing the dipole-dipole
interaction we discuss the effects of entanglement in this
many-body system, showing that the critical temperature of the
transition  from the fluorescent to super-radiant phase is not
modified in this case. We proved that the spectrum of the
collective bosonic excitations of the model is unaffected by the
dipole-dipole interaction. In this case, there is also a quantum
phase transition, at some values of the physical parameters of the
model. For the reader interested in the the study of entanglement
in many-body systems close to the quantum phase transition, see
for example the Refs. \cite{osborn} \cite{amico1} \cite{kitaev}.

Recently, using the path integral approach with functional
integration method the analytic behavior of thermodynamic
quantities in the full Dicke model \cite{dicke} \cite{hl1}
\cite{wanghioe} \cite{hl2} \cite{hioe} \cite{duncan} was presented
\cite{tese}. See also the Refs. \cite{pizi} \cite{pizi2}. The full
Dicke model is similar to the Jaynes-Cummings model \cite{jc}
where the non-resonant processes in which the atom and the field
are excited or de-excited simultaneously, known in the literature
as the anti-Jaynes-Cummings model. In Ref. \cite{tese} the study
of the nonanalytic behavior of thermodynamic quantities in the
full Dicke model, allow the authors to evaluate the critical
transition temperature. It was shown that the system develop a
quantum phase transition and a phase transition from ordinary
fluorescence to super-radiant phase at some critical temperature.
Based in this analysis, the spectrum of the collective bosonic
excitations, for different situations were presented. It was study
the rotating-wave approximation, the counter rotating-wave
approximation and also the general case. As we discussed, the
rotating-wave approximation ignores energy non-conserving terms in
which the emission (absorption) of a quantum of a quantized field
is accompanied by the transition of one atom from its lower
(upper) to its upper (lower) state. In a situation where only
non-resonant processes contribute, the full Dicke model present a
second order phase transition from the ordinary fluorescent to the
super-radiant phase respectively, at some critical temperature
$\beta^{-1}_{c}$ and also a quantum phase transition, i.e., a
phase transition at zero temperature. In the last case there are
no thermal excitations, therefore the phase transition is driven
by the quantum fluctuations. The interesting result is the fact
that it is possible to have a condensate with super-radiance in a
system of $N$ two-level atoms coupled with one mode of a bosonic
field where only non-resonant processes contribute.

In the Ref. \cite{artigo}, still neglecting the direct interaction
between the atoms, it was considered also two different models,
assuming that a single quantized mode of a bosonic field interacts
with a ensemble of $N$ identical two-level atoms. Again we assume
that the system is in thermal equilibrium with a reservoir at
temperature $\beta^{-1}$. Analytic properties of the partition
functions of the models were also investigated. First, it was
study a modified version of the model discussed by Chang and
Chakravarty, Legget and others \cite{chang} \cite{legget}
\cite{parma} \cite{benatti}, which has been used to analyze
dissipation in quantum computers. Owing to the coupling between
the two-level systems and a bosonic reservoir, this model presents
destruction of quantum coherence without decay of population.
Since the interaction Hamiltonian of the model generates a quite
particular non-resonant processes, it is shown that the partition
function is analytic for all temperatures, and therefore there is
no second order phase transition in the model. Second, it was
investigated a model where the coupling between the bosonic mode
and $N$ two-level atoms is intensity dependent \cite{jj1}
\cite{jj2} \cite{jj3}, introducing also the couplings $g_{1}$ and
$g_{2}$ for rotating and counter-rotating terms respectively. At
low temperatures, the contribution coming from the
counter-rotating terms dominates over the rotating ones. When the
coupling constant is given by
$g_{2}=(\omega_{0}\,\Omega)^{\frac{1}{2}}$, where $\omega_{0}$ is
the energy of the single-mode bosonic field and $\Omega$ is the
energy gap of the atoms, a quantum phase transition appears.

In this paper the physical system in consideration is an ensemble
of two-level atoms interacting with a single mode of a bosonic
field. It is a natural question to ask if non-Gaussian terms can
change the critical behavior of this spin-boson model. We study
two models, the full Dicke model and the same model where we
include the dipole-dipole interaction term between the two-level
atoms. We proved that the critical temperature that characterize
the phase transition from fluorescent to super-radiant phase is
the same in both models. Also the quantum phase transitions, at
some values of the physical parameters of both models are the
same. We are using the path integral approach with the functional
integration method to investigate the thermodynamic of the models,
which is given by the analytic properties of the partition
function. The paper is organized as follows. In section II we
present the fermionic full Dicke model and the model with the
dipole-dipole interaction. In section III we discuss the
functional integral for the full Dicke model. In section IV we
study the full Dicke model with the dipole-dipole interaction.
Conclusion are given in section V. In the Appendix the theory of
pure and mixed ensembles and the reduced density operator is
briefly presented. In the paper we use $k_{B}=c=\hbar=1$.

\section{The $N$ two-level atoms-Bose field interaction Hamiltonians}\

In this section we consider a very general situation where the
system under investigation contains a large number of two-level
atoms. In order to describe the dynamics of the the two-level
atoms and the bosonic mode we have to introduce the Hamiltonian
governing the interaction of the quantized Bose field with free
atoms \cite{puri} \cite{ja2}. Free means that there is no
interaction between the atoms. Therefore let us consider a Bose
quantum system $B$, with Hilbert space ${\cal H}^{(B)}$ which is
coupled with $N$ atoms, with Hilbert space  ${\cal H}^{(Q)}$. Let
us assume that the whole system is in thermal equilibrium at
temperature $\beta^{-1}$. The Bose quantum system is a sub-system
of the total system living in the tensor product space ${\cal
H}^{(B)}\,\otimes\,{\cal H}^{(Q)}$. Let us denote by $H_{B}$ the
Hamiltonian of the quantized Bose field, by $H_{Q}$  the free
Hamiltonian of the $N$ two-level atoms and $H_{I}$ the Hamiltonian
describing the interaction between the quantized Bose field and
the $N$ two-level atoms. The Hamiltonian for the total system can
be written as
\begin{equation}
H=H_{B}\,\otimes\,I_{Q}+I_{B}\,\otimes\,H_{Q}+\delta H_{I},
\label{13}
\end{equation}
where $I_{B}$ and $I_{Q}$ denotes the identities in the Hilbert
spaces of the quantized Bose field and the $N$ atoms. In the Eq.
(\ref{13}), $\delta$ is a small coupling constant between the
atoms and the quantized Bose field.

The main purpose of this section is to discuss $N$-atoms-Bose
field interaction Hamiltonian. Therefore, let us introduce the
Dicke operators to describe each atom. The free $j-th$ atom
Hamiltonian will be denoted by $H_{D}^{(j)}$, since we are using
the Dicke representation. Therefore, we have
\begin{equation}
H_{D}^{(j)}|\,i\,\rangle_{j}=\omega_{i}^{(j)}|\,i\,\rangle_{j},
\label{14}
\end{equation}
where $|\,i\,\rangle_{j}$ are orthogonal energy eigenstates
accessible to the $j-th$ atom and $\omega_{i}^{(j)}$ are the
respective eigenfrequencies. Using Eq. (\ref{14}) and the
orthonormality of the energy eigenstates we can write the  $j-th$
atom Hamiltonian $H_{D}^{(j)}$ as
\begin{equation}
H_{D}^{(j)}=\sum_{i=1}^{2}\,\omega_{i}^{(j)} ( |\,i\,\rangle \,
\langle\, i| \, )_j . \label{15}
\end{equation}
Let us define the pseudo-spin operators $\sigma_{(j)}^z$,
$\sigma_{(j)}^{+}$ and $\sigma_{(j)}^{-}$ for each atom by
\begin{equation}
\sigma_{(j)}^z=\frac{1}{2} \, ( |2\,\rangle \, \langle\,2|-
\,|1\,\rangle \, \langle\,1| \, ) _j , \label{16}
\end{equation}
\begin{equation}
\sigma_{(j)}^{+}=\,(|2\,\rangle \, \langle\,1| \, )_j , \label{17}
\end{equation}
and finally
\begin{equation}
\sigma_{(j)}^{-}=\, (|1\,\rangle \, \langle\,2| \, )_j .
\label{18}
\end{equation}
Combining Eq. (\ref{15}) and Eq. (\ref{16}), the $j-th$ atom
Hamiltonian can be written as
\begin{equation}
H_{D}^{(j)}=\,\Omega^{(j)}\,\sigma_{(j)}^z+\frac{1}{2}
\biggl(\omega_{1}^{(j)}+\omega_{2}^{(j)}\biggr), \label{19}
\end{equation}
where the energy gap between the energy eigenstates of the $j-th$
atom is given by
\begin{equation}
\Omega^{(j)}=\omega_{2}^{(j)}-\omega_{1}^{(j)}.
\label{20}
\end{equation}
Shifting the zero of energy to
$\frac{1}{2}(\omega_{1}^{(j)}+\omega_{2}^{(j)})$ for each atom,
the $j-th$ atomic Hamiltonian given by Eq. (\ref{19}) can be
rewritten as
\begin{equation}
H_{D}^{(j)}=\Omega^{(j)}\,\sigma_{(j)}^z.
\label{21}
\end{equation}
Note that the pseudo-spin operators $\sigma_{(j)}^{+}$,
$\sigma_{(j)}^{-}$ and $\sigma_{(j)}^z$ satisfy the standard
angular momentum commutation relations corresponding to spin
$\frac{1}{2}$ operators, i.e.,
\begin{equation}
\left[\sigma_{(j)}^{+},\sigma_{(j)}^{-}\right]=2\,\sigma_{(j)}^z,
\label{22}
\end{equation}
\begin{equation}
\left[\sigma_{(j)}^z,\sigma_{(j)}^{+}\right]=\sigma_{(j)}^{+},
\label{23}
\end{equation}
and finally
\begin{equation}
\left[\sigma_{(j)}^z,\sigma_{(j)}^{-}\right]=-\sigma_{(j)}^{-}.
\label{24}
\end{equation}
The model that we are interested is composed by the Hamiltonian of
the atoms, with the contribution of the quantized Bose field
$H_{S}$, and the interaction Hamiltonian $H_{I}^{(j)}$. The
Hamiltonian of the total system is given by
\begin{eqnarray}
&&
H= I_B\,\otimes\, H_{Q}+H_{B}\,\otimes I_Q+\delta H_{I}=\nonumber\\
&&I_B\,\otimes\,\Omega\sum_{j=1}^{\infty}
\,\sigma_{(j)}^{z}+\sum_{k=1}^{\infty}
\omega_{k}\,b_{k}^{\dagger}\,b_{k}\,\otimes\,I_Q+
\frac{1}{\sqrt{N}}\sum_{j=1}^{N}\sum_{k=1}^{\infty}
 g_{kj}\,\left(
b_{k}+b_{k}^{\dagger}\right)\,\otimes \left(\sigma_{(j)}^{+}+
\sigma_{(j)}^{-}\right). \label{32}
\end{eqnarray}
where the first summation in the right hand side is
\begin{equation}
\sum_{j=1}^{N}\, \Omega^{(j)}\,\sigma_{(j)}^z=
\Omega^{(1)}\,\sigma_{(1)}^z\,\otimes{\bf{1}}\,
\otimes...\otimes{\bf{1}}+...+
{\bf{1}}\,\otimes{\bf{1}}\otimes...\otimes{\bf{1}}
\otimes\,\Omega^{(N)}\,\sigma_{(N)}^z,
\label{def}
\end{equation}
and $\bf{1}$ denotes the identity in the Hilbert space of each
two-level system. The quantity $g_{kj}$ is the coupling
coefficient. Each two-level system is described by a Hilbert space
${\cal H}^{(1)},{\cal H}^{(2)},...,{\cal H}^{(N)}$. The composite
atomic system is described by the $N$-fold tensor product space
${\cal H}^{(1)}\otimes {\cal H}^{(2)}\otimes {\cal H}^{(3)}..$.

One point which is important to stress is that in the Eq.
(\ref{32}) there are terms so called counter-rotating terms.  In
the rotating-wave-approximation we ignore energy non-conserving
terms in which the emission (absorption) of a quantum of a
quantized field is accompanied by the transition of one atom from
its lower (upper) to its upper (lower) state. As we discussed
before, the rotating-wave-approximation ignores terms in which the
$j-th$ atomic raising (lowering) operators multiplies the field
creation (annihilation) operator.

We can also introduce an atom-atom interaction, which is relevant
in the study of entangled states. In an entangled system, the
state of the composite system can not be factorized in to a
product of the states of its sub-systems. For example in the case
of two two-level atoms, the interaction Hamiltonian of two atoms
including the dipole-dipole interaction reads
\begin{equation}
H_{(qq)}=\frac{1}{N}\sum_{i\neq
j}^{2}H_{(ij)}\,\sigma_{(i)}^{+}\otimes\sigma_{(j)}^{-}.
\label{dipole}
\end{equation}
Let us note that the interaction Hamiltonian given by Eq.
(\ref{dipole}) is a particular case of the quantum Hamiltonian
describing an axially symmetric ferromagnet, with the coupling
constants $J_{ij}^{z}$ and $J_{ij}$. In terms of the pseudo spin
operators we have
\begin{equation}
H=-\sum_{<i,j>}\,\Bigg(J_{ij}^{z}\,\sigma_{(i)}^{z}\,\sigma_{(j)}^{z}+
J_{ij}\biggl(\sigma_{(i)}^{+}\sigma_{(j)}^{-}+\sigma_
{(i)}^{-}\sigma_{(j)}^{+}\biggr)\Biggr). \label{dipole2}
\end{equation}
The $<i,j>$ symbol denotes a summation over pair of first
neighbors. In the absence of the "dipole-dipole" interaction the
pure Hilbert space of the two atomic system is spanned by the
states $| \, g_{1} \, \rangle\,\otimes\,| \, g_{2} \, \rangle$, $|
\, g_{1} \, \rangle\,\otimes\,| \, e_{2} \, \rangle$, $| \, e_{1}
\, \rangle\,\otimes\,| \, g_{2} \, \rangle$ and $| \, e_{1} \,
\rangle\,\otimes\,| \, e_{2} \, \rangle$, where $g$ and $e$
denotes respectively the ground and the excited state of the two
atomic system. If we include the dipole-dipole interaction term in
the form of Eq. (\ref{dipole}), the vectors $| \, g_{1} \,
\rangle\,\otimes\,| \, e_{2} \, \rangle$ and  $| \, e_{1} \,
\rangle\,\otimes\,| \, g_{2} \, \rangle$ are not more eigenstates
of the Hamiltonian of the atomic system. It can be shown that
these two vectors states must be substituted by the two entangled
states, known in the literature as maximally entangled states
\cite{mann} \cite{tanas}
\begin{equation}
| \, s \,\rangle=\frac{1}{\sqrt{2}}\,\left( | \, e_{1}
\,\rangle\,\otimes\,| \, g_{2} \, \rangle+ | \, g_{1}
\,\rangle\,\otimes\,| \, e_{2} \, \rangle\right)
 \label{dipole2}
\end{equation}
and
\begin{equation}
| \, a \,\rangle=\frac{1}{\sqrt{2}}\,\left( | \, e_{1}
\,\rangle\,\otimes\,| \, g_{2} \, \rangle- | \, g_{1}
\,\rangle\,\otimes\,| \, e_{2} \, \rangle\right).
 \label{dipole3}
\end{equation}
The Hilbert space of the system is spanned by the collective
states of the two identical interacting two-level atoms given
respectively by $(| \, g_{1} \, \rangle\,\otimes\,| \, g_{2} \,
\rangle, | \, s \, \rangle, | \, a \, \rangle,| \, e_{1} \,
\rangle\,\otimes\,| \, e_{2} \, \rangle)$. As stated before, both
the states $| \, s \,\rangle$ and $| \, a \,\rangle$ tell us
nothing about the state of the first or second atom separately.
Therefore when we include the dipole-dipole interaction between
the atoms, we open the possibility of realize entangled states.

In the spin-boson model with the dipole-dipole interaction
included, we are considering the question of how does the
dipole-dipole interaction change the critical temperature, where
again the system exhibit a phase transition from fluorescence to
super-radiance. The Hamiltonian for the spin-boson model with the
dipole-dipole interaction reads
\begin{eqnarray}
&& H= \frac{1}{N}\sum_{i\neq
j}^{N}H_{(ij)}\,\sigma_{(i)}^{+}\otimes\sigma_{(j)}^{-}+ \nonumber\\
&&I_B\,\otimes\,\Omega\sum_{j=1}^{N}
\,\sigma_{(j)}^{z}+\sum_{k=1}^{\infty}
\omega_{k}\,b_{k}^{\dagger}\,b_{k}\,\otimes\,I_Q+
\frac{1}{\sqrt{N}}\sum_{j=1}^{N}\sum_{k=1}^{\infty}
 g_{ik}\left(
b_{k}+b_{k}^{\dagger}\right)\,\otimes \left(\sigma_{(j)}^{+}+
\sigma_{(j)}^{-}\right). \label{aaa}
\end{eqnarray}
Each two-level atom interact with all other atoms of the ensemble
with the same coupling strength, i.e., $H_{(ij)}=\lambda$, and the
summation is over all the atoms. This is a model with one infinite
range interaction. This model is based in the models describing
spin systems with long range interaction \cite{haldane}
\cite{pittel}. In the next section we discuss first the full Dicke
model, showing the presence of a quantum phase transition and also
a super-radiant phase at some critical temperature.

\section{The functional integral for the fermionic full Dicke model}

\quad $\,\,$ The typical situation in cavity quantum
electrodynamics is the case where a single atom is coupled to a
cavity mode. Now suppose $N$ identical two-level atoms in an
optical cavity whose linear dimensions are small compared to the
wavelength associated to the bosonic mode. This ensemble of
identical $N$ two-level atoms interacting linearly with one mode
of a bosonic field defines the full Dicke model. The Hamiltonian
of the system reads
\begin{equation}
H= I_S\,\otimes\,\sum_{j=1}^{N}\,
\frac{\Omega}{2}\,\sigma_{(j)}^z+\omega_{0}\,b^{\dagger}\,b\,\otimes\,I_B+
\,\frac{g}{\sqrt{N}} \sum_{j=1}^{N}\,
\Bigl(b+b^{\dagger}\Bigr)\otimes
\Bigl(\sigma_{(j)}^{+}+\sigma_{(j)}^{-}\Bigr)\, . \label{i26}
\end{equation}
In the above equation $g$ is the coupling constant between the
atoms and the single mode of the bosonic field. The $b$ and
$b^{\dagger}$ are the boson annihilation and creation operators of
mode excitations that satisfy the usual commutation relation
rules.

The aim of this section is to prove that a model with an
interaction Hamiltonian where we take into account non-resonant
processes in which the two-level systems and the bosonic modes are
excited or de-excited simultaneously, presents a phase transition
from normal to super-radiant state at some temperature with the
presence of a condensate and also a quantum phase transition at
some critical coupling. We should mention that the results of this
section are not original, but by completeness we include in the
paper. The Euclidean action of the model is $S$, where $H_{F}$ is
the full Hamiltonian for the fermionic full Dicke model. This
quantity can be written as
\begin{eqnarray}
H_{F}\,=\,\omega_{0}\,b^{\,*}(\tau)\,b(\tau)\,+
\,\frac{\Omega}{2}\,\displaystyle\sum_{i\,=\,1}^{N}\,
\biggl(\alpha^{\,*}_{\,i}(\tau)\,\alpha_{\,i}(\tau)\,-
\,\beta^{\,*}_{\,i}(\tau)\beta_{\,i}(\tau)\biggr)\,+
\nonumber\\
+\,\frac{g_{\,1}}{\sqrt{N}}\,\displaystyle\sum_{i\,=\,1}^{N}\,
\biggl(\alpha^{\,*}_{\,i}(\tau)\,\beta_{\,i}(\tau)\,b(\tau)\,+
\alpha_{\,i}(\tau)\,\beta^{\,*}_{\,i}(\tau)\,b^{\,*}(\tau)\,\biggr)\,+
\nonumber\\
+\,\frac{g_{\,2}}{\sqrt{N}}\,\displaystyle\sum_{i\,=\,1}^{N}\,
\biggl(\alpha_{\,i}(\tau)\,\beta^{\,*}_{\,i}(\tau)\,b(\tau)\,+
\,\alpha^{\,*}_{\,i}(\tau)\,\beta_{\,i}(\tau)\,b^{\,*}(\tau)\biggr).
\label{66a}
\end{eqnarray}
Note we are introducing two coupling constants, $g_{1}$ and
$g_{2}$, for the rotating and counter-rotating terms,
respectively. As we discussed before, the main reason for this is
that we are interested in to identify the contribution of the real
and virtual processes in the phase transition with the formation
of the condensate. For this purpose we must to find the partition
function of the system, therefore we calculate the formal quotient
$\frac{Z}{Z_{0}}$, being defined by the equation:
\begin{equation}
\frac{Z}{Z_{0}}=\frac{\int [d\eta]\,e^{\,S}}{\int
[d\eta]\,e^{\,S_{0}}}\, , \label{partition1}
\end{equation}
where the functional integral is with respect to the complex
functions $b^*(\tau)$ and $b(\tau)$ and Grassmann Fermi fields
$\alpha_i^*(\tau)$, $\alpha_i(\tau)$, $\beta_i^*(\tau)$ and
$\beta_i(\tau)$. Here the expression $[d\eta]$ is the functional
measure. It is possible to show that an expression for
$\frac{Z}{Z_{\,0}}$ which is given by \cite{tese} \cite{artigo}
\begin{eqnarray}
\frac{Z}{Z_{\,0}}&=& \Biggl[\,\biggl(1\,-\,a(0)\,+
\,2\,c(0)\biggr)\,\biggl(1\,-\,a(0)\,-\,2\,c(0)\,\biggr)\,\Biggr]^{\,-\,1/2}\,
\nonumber\\
&&\prod_{\,\omega\,>\,0}\,\Biggl[\,\biggl(1\,-\,a(\omega)\,\biggr)\,
\biggl(1\,-\,a(\,-\omega)\,\biggr)\,-\,4c^{\,2}(\omega)\,\Biggr]^{\,-\,1}\,+
\nonumber
\\
\nonumber
\\
&&+O(N^{\,-1})\,,
 \label{95}
\end{eqnarray}
where $a(\omega)$ and $c(\omega)$ in the above equation are given,
respectively, by
\begin{equation}
a(\omega)\,=\,\Biggl(\frac{g_{\,1}^{\,2}\,(\Omega
-i\omega)^{\,-1}+\,g_{\,2}^{\,2}\,(\Omega\,+\,
i\omega)^{\,-1}}{(\omega_{0}\,-\,i\,\omega)}
\Biggr)\,\tanh{\biggl(\,\frac{\beta\,\Omega}{4}\,\biggr)}\,
\label{90}
\end{equation}
and
\begin{equation}
c(\omega)\,=\,\Biggl(\frac{g_{\,1}\,g_{\,2}\,\Omega}{(\omega_{\,0}^{\,2}\,+
\,\omega^{\,2})^{\,1/2}\,(\Omega^{\,2}\,+\,\omega^{\,2})}\Biggr)
\,\tanh{\,\biggl(\frac{\beta\,\Omega}{4}\biggr)}.
 \label{90a}
\end{equation}
To obtain the thermodynamic limit we have to take the limit
($N\rightarrow \infty$) in Eq. (\ref{95}). We turn out to the
discussion concerning the local elementary excitation of the
ground state. To find the energy spectrum of the bosonic
collective excitations we have to use the equation
\begin{equation}
4\,c^{\,2}(\omega)\,-\,\biggl(1\,-\,a(\omega)\,\biggr)\,
\biggl(1\,-\,a(\,-\omega)\,\biggr)\,=0\, , \label{105}
\end{equation}
and making the analytic continuation $(i\omega \rightarrow E)$, we
obtain the following equation
\begin{eqnarray}
&&1\,=\,-\Biggl[\frac{g_{\,1}^{\,4}\,+\,g_{\,2}^{\,4}}
{(\omega_{\,0}^{\,2}\,-\,E^{\,2})\,(\Omega^{\,2}\,-\,E^{\,2})}\Biggr]\,
\tanh^{\,2}\biggl(\frac{\beta\,\Omega}{4}\biggr)\,+
\nonumber\\
\nonumber\\
&&-\Biggl[\frac{g_{\,1}^{\,2}\,g_{\,2}^{\,2}}{(\omega_{\,0}^{\,2}
\,-\,E^{\,2})}\Biggl(\frac{1}{(\Omega\,-E)^{\,2}}\,+
\,\frac{1}{(\Omega\,+\,E)^{\,2}}\,-\,\frac{4\,\Omega^{\,2}}
{(\Omega^{\,2}\,-\,E^{\,2})^{\,2}}\Biggr)\,\Biggr]\,
\tanh^{\,2}\Biggl(\frac{\beta\,\Omega}{4}\Biggr)\,+
\nonumber\\
\nonumber\\
&&+\,\Biggl[\frac{g_{\,1}^{\,2}(\,\Omega\,-\,E\,)^{\,-1}\,+
\,g_{\,2}^{\,2}(\,\Omega\,+\,E\,)^{\,-1}}{(\omega_{0}\,-\,E)}\,+
\frac{g_{\,1}^{\,2}(\,\Omega\,+\,E\,)^{\,-1}\,+
\,g_{\,2}^{\,2}(\,\Omega\,-\,E\,)^{\,-1}}{(\omega_{0}\,+\,E)}\Biggr]
\,\tanh\Biggl(\frac{\beta\,\Omega}{4}\Biggr).\nonumber\\
\label{105}
\end{eqnarray}
Solving the above equation for the case $\beta^{-1}=\beta^{-1}_c$
we find the following roots
\begin{equation}
E_{\,1}\,=\,0
\label{106}
\end{equation}
and
\begin{equation}
E_{\,2}\,=\,\Biggl(\,\frac{g_{\,1}\,(\Omega\,+\,\omega_{\,0})^{\,2}\,+\,
g_{\,2}\,(\Omega\,-\,\omega_{\,0})^{\,2}}{(g_{\,1}\,+\,g_{\,2})}\,\Biggr)^{\,1/2}\,.
\label{107}
\end{equation}
Its low energy state of excitation is a Nambu-Goldstone mode,
since the continuous $U(1)$ symmetry has been broken. Therefore by
using the non-relativistic version for the Goldstone theorem, the
energy spectrum can not have a gap above the ground state. Now,
let us present the critical temperature and the energy level
spectrum of the collective bosonic excitations of the model with
the rotating-wave approximation, where $g_{1}\neq 0$ and
$g_{2}=0$.
The result obtained by Popov and Fedotov \cite{popov1}
\cite{popov2} is recovered, where the equation
\begin{equation}
a(0) = 1\, \label{102}
\end{equation}
and
\begin{equation}
\frac{g_{1}^{2}}{\omega_{0}\Omega}
\tanh\biggl(\frac{\beta_{c}\,\Omega}{4}\biggr) = 1\,, \label{103}
\end{equation}
gives the inverse of the critical temperature, $\beta_{c}$. It is
given by
\begin{equation}
\beta_{c} = \frac{4}{\Omega}\,arctanh
\biggl(\frac{\omega_{0}\Omega}{g_{1}^{2}}\biggr)\, . \label{104}
\end{equation}
The order parameter of the transition is the expectation value of
the number of excitation associated to the bosonic mode per atom,
i.e., $lim_{N\,\rightarrow\infty}\,\frac{\langle\,b^
{\dagger}\,b\rangle}{N}\neq\,0$. Note that again $\omega_{0}$,
$\Omega$ and $g_{1}$ define also a non-zero critical temperature
where the partition function is no more analytic. We may expect a
super-radiant phase for the temperature $\beta_{c}^{-1}$ given by
Eq. (\ref{104}). The energy level spectrum of the collective Bose
excitations in this case is
\begin{equation}
E_{1}=0\, , \label{108}
\end{equation}
and
\begin{equation}
E_{2}=\Omega+\omega_{0}\, . \label{109}
\end{equation}
In this case, there is also a quantum phase transition, i.e., a
zero temperature phase transition when the coupling constant
$g_{1}$ attains the value
$g_{1}=(\omega_{0}\,\Omega)^{\frac{1}{\,2}}$. Now we will show
that is possible to have a condensate with super-radiance in a
system of $N$ two-level atoms coupled with one mode of a Bose
field where only virtual processes contribute.
In the pure counter-rotating wave case, i.e., $g_{1}=0$ and
$g_{2}\neq 0$, the inverse of the critical temperature,
$\beta_{c}$ is given by
\begin{equation}
\beta_{c} = \frac{4}{\Omega}\,arctanh
\biggl(\frac{\omega_{0}\Omega}{g_{2}^{2}}\biggr)\, , \label{112}
\end{equation}
and the spectrum of the collective Bose excitations given by
\begin{equation}
E_{1}=0\, , \label{110}
\end{equation}
and
\begin{equation}
E_{2}=|\,\Omega-\omega_{0}|\, . \label{111}
\end{equation}
Again, there is a zero temperature phase transition when
$g_{2}=(\omega_{0}\,\Omega)^{\frac{1}{2}}$. In this case there is
no thermal excitations and the phase transition is driven by the
quantum fluctuations. A comment is in order concerning the
spectrum of the Bose excitations. In both of the cases: working
with the pure counter-rotating or the rotating-wave terms, there
is a phase transition. In the case of the rotating-wave
approximation $g_{1}\neq 0$ and $g_{2}=0$, there is a
Nambu-Goldstone mode $(E=0)$. In the pure counter-rotating case
$g_{1}=0$ and $g_{2}\neq 0$ also there is a Nambu-Goldstone
(gapless) mode. We show that it is possible to have a condensate
with super-radiance in a system of $N$ two-level atoms coupled
with one mode of a Bose field where only virtual processes
contribute. Since the energy $E_{2}\approx 0$, local elementary
excitations of the ground state with low energy can easily be
created causing a significant fluctuation effect. Unfortunately we
are not able to evaluate these effect in the systems.

\section{The functional integral for the spin-boson
model with dipole-dipole coupling}

\quad $\,\,$ The aim of this section is to prove that the
introduction of the dipole-dipole interaction does not modify the
temperature of the phase transition from normal to super-radiant
state and the presence of a condensate. Also, the energy spectrum
of the collective bosonic excitations of the model is the same as
in the full Dicke model. Introducing the dipole-dipole coupling in
the full Dicke model, where an ensemble of identical $N$ atoms
interacts linearly with one mode of a bosonic field, in this
section we use the same approach of functional integrals to study
the thermodynamic of the model \cite{tese} \cite{artigo}
\cite{popov1} \cite{popov2}. We would like to stress that in this
new situation, the Grassmann functional integrals are not
Gaussian. In order to perform the integrals we introduce auxiliary
variables in the calculation. The total Hamiltonian of this
spin-boson model is defined by:
\begin{eqnarray}
&& H_{T}= \frac{1}{N}\sum_{i\neq
j}^{N}H_{(ij)}\,\sigma_{(i)}^{+}\otimes\sigma_{(j)}^{-}+ \nonumber\\
&&I_B\,\otimes\,\Omega\sum_{j=1}^{N} \,\sigma_{(j)}^{z}+
\omega_{0}\,b^{\dagger}\,b\,\otimes\,I_Q+
\frac{g}{\sqrt{N}}\sum_{j=1}^{N}
 \left(
b+b^{\dagger}\right)\,\otimes \left(\sigma_{(j)}^{+}+
\sigma_{(j)}^{-}\right). \label{b31}
\end{eqnarray}
In the above equation the quantity $g$ is the coupling constant
between the atoms and the single mode of the bosonic field. The
$b$ and $b^{\dagger}$ are the boson annihilation and creation
operators of mode excitations that satisfy the usual commutation
relation rules. As usual, we are using the pseudo-spin operators
$\sigma_{(j)}^{+}$, $\sigma_{(j)}^{-}$ and $\sigma_{(j)}^z$ which
satisfy the standard angular momentum commutation relations
corresponding to spin $\frac{1}{2}$ operators. We are also
shifting the zero of energy $\frac{1}{2}(\omega_{1}+\omega_{2})$
for each atom and defining $\Omega=\omega_{2}-\omega_{1}$.

Let us define the Fermi raising and lowering operators
$\alpha^{\dagger}_{i}$, $\alpha_{i}$, $\beta^{\dagger}_{i}$ and
$\beta_{i}$, that satisfy the anti-commutator relations
$\alpha_{i}\alpha^{\dagger}_{j}+\alpha^{\dagger}_{j}\alpha_{i}
=\delta_{ij}$ and
$\beta_{i}\beta^{\dagger}_{j}+\beta^{\dagger}_{j}\beta_{i}
=\delta_{ij}$. We can also define the following bilinear
combination of Fermi operators, $\alpha^{\dagger}_{i}\alpha_{i}
-\beta^{\dagger}_{i}\beta_{i}$, $\alpha^{\dagger}_{i}\beta_{i}$
and finally $\beta^{\dagger}_{i}\alpha_{i}$. Note that
$\sigma^z_{(\,i)}$, $\sigma^+_{(\,i)}$ and $\sigma^-_{(\,i)}$ obey
the same commutation relations as the above bilinear combination
of Fermi operators. Therefore, we can change the pseudo-spin
operators of this spin-boson model by using the bilinear
combination of Grassmann Fermi fields

\begin{equation}
\sigma_{(i)}^{z}\longrightarrow (\alpha_{i}^{\dagger}\alpha_{i}
-\beta_{i}^{\dagger}\beta_{i})\, , \label{34}
\end{equation}
\begin{equation}
\sigma_{(i)}^{+}\longrightarrow \alpha_{i}^{\dagger}\beta_{i}\, ,
\label{35}
\end{equation}
and finally
\begin{equation}
\sigma_{(i)}^{-}\longrightarrow \beta_{i}^{\dagger}\alpha_{i}\, .
\label{36}
\end{equation}
Using the Eq. (\ref{b31}), the Euclidean fermionic action for the
model can be written as
\begin{equation}
S = \int_0^{\beta} d\tau \left( b^*(\tau) \frac{\partial
b(\tau)}{\partial \tau} + \sum_{i=1}^{N} \left( \alpha_i^*(\tau)
\frac{\partial \alpha_i(\tau)}{\partial \tau} + \beta_i^*(\tau)
\frac{\partial \beta_i(\tau)}{\partial \tau} \right) \right) -
\int_0^{\beta} d\tau H_F(\tau)\,. \label{s1}
\end{equation}
The Euclidean action $S$ is given by Eq. (\ref{s1}) where $H_{F}$
is the full Hamiltonian for the fermionic full Dicke model with
dipole-dipole interaction. As in the previous section, introducing
two coupling constants, $g_{1}$ and $g_{2}$, for the rotating and
counter-rotating terms, the Hamiltonian becomes
\begin{eqnarray}
H_{F}&=&\omega_{0}\,b^{\,*}(\tau)\,b(\tau)\,+
\,\frac{\Omega}{2}\,\displaystyle\sum_{i\,=\,1}^{N}\,
\biggl(\alpha^{\,*}_{\,i}(\tau)\,\alpha_{\,i}(\tau)\,-
\,\beta^{\,*}_{\,i}(\tau)\beta_{\,i}(\tau)\biggr)\,+
\nonumber\\
&+&\frac{1}{N}
\sum_{i\neq
j}^{N}H_{(ij)}\,\alpha^{\,*}_{\,i}(\tau)\,\beta_{\,i}(\tau)\,
\beta^{\,*}_{\,j}(\tau)\alpha_{\,j}(\tau)+\frac{g_{\,1}}{\sqrt{N}}
\,\displaystyle\sum_{i\,=\,1}^{N}\,
\biggl(\alpha^{\,*}_{\,i}(\tau)\,\beta_{\,i}(\tau)\,b(\tau)\,+
\alpha_{\,i}(\tau)\,\beta^{\,*}_{\,i}(\tau)\,b^{\,*}(\tau)\,\biggr)\,+
\nonumber\\
&+&\frac{g_{\,2}}{\sqrt{N}}\,\displaystyle\sum_{i\,=\,1}^{N}\,
\biggl(\alpha_{\,i}(\tau)\,\beta^{\,*}_{\,i}(\tau)\,b(\tau)\,+
\,\alpha^{\,*}_{\,i}(\tau)\,\beta_{\,i}(\tau)\,b^{\,*}(\tau)\biggr).
\label{b32}
\end{eqnarray}
Let us define the formal quotient of two functional integrals,
i.e., the partition function of the interacting model and the
partition function of the free model. Therefore, we are interested
in to calculate the quantity given by Eq. (\ref{partition1}) where
$S=S(b,b^*,\alpha,\alpha^{\dagger},\beta,\beta^{\dagger})$ is the
Euclidean action given by Eq. (\ref{s1}). Let us define
$S_0=S_{0}(b,b^*,\alpha,\alpha^{\dagger},\beta,\beta^{\dagger})$
which is the free Euclidean action for the free single bosonic
mode and the free two-level atoms and finally $[d\eta]$ is the
path integral measure. Note that in Eq. (\ref{partition1}) we have
functional integrals with respect to the complex functions
$b^*(\tau)$ and $b(\tau)$ and Grassmann Fermi fields
$\alpha_i^*(\tau)$, $\alpha_i(\tau)$, $\beta_i^*(\tau)$ and
$\beta_i(\tau)$. Since we are using thermal equilibrium boundary
conditions in the imaginary time formalism, the integration
variables in Eq. (\ref{partition1}) obey periodic boundary
conditions for the Bose field, i.e., $b(\beta)=b(0)$, and
anti-periodic boundary conditions for the Grassmann Fermi fields
i.e., $\alpha_i(\beta)=-\alpha_i(0)$ and $
\beta_i(\beta)=-\beta_i(0)$.
The free action of the bosonic field is given by
\begin{equation}
S_{0}(b,b^*) = \int_{0}^{\beta} d\tau \biggl(b^{*}(\tau)
\frac{\partial b(\tau)}{\partial \tau} -
\omega_{0}\,b^{*}(\tau)\,b(\tau)\biggr)\,.
\end{equation}
In order to obtain the effective action of the bosonic mode we
must integrate over the Grassmann Fermi fields in the expression
given by Eq. (\ref{partition1}). The problem now confronting us is
the fact that in the action given by Eq. (\ref{b32}) there is a
non-Gaussian term, corresponding to the dipole-dipole interaction.
Although it is not possible to integrate this term directly, using
auxiliary variables of integration we can circumvented this
difficult. The following integral is helpful
\begin{eqnarray}
\exp{\left(-\frac{1}{N}\sum_{i,\,j=1}^N x^*_iA_{ij}x_j\right)}=(\det{A})^{-1}
\int\prod_{i=1}^N\frac{dy^*_idy_i}{-2\pi i}
e^{\sum_{i,\,j=1}^N y^*_iA^{-1}_{ij}y_j+\frac{1}{\sqrt{N}}\sum_{i=1}^N
y_ix^*_i+\frac{1}{\sqrt{N}}\sum_{i=1}^N y^*_ix_i}\,,
\label{intreal}
\end{eqnarray}
where this integral is valid for commutating variables. To
continue we have to study the non-Gaussian contribution to the
partition function. In terms of the Grassmann variables we define
$R$ as
\begin{eqnarray}
R&=&\exp{\left(-\frac{1}{N}\sum_{i,\,j=1}^N
\int_0^{\beta}d\tau\,\alpha^*_i(\tau)\beta_i
(\tau)\,H_{ij}\,\beta^*_j(\tau)\alpha_j(\tau)\right)}\nonumber\\
&=&\exp{\left(-\frac{1}{N}\sum_{i,\,j=1}^N \int_0^{\beta}d\tau\,
\Bigl(\beta^*_i(\tau)\alpha_i(\tau)\Bigr)^*\,H_{ij}\,
\beta^*_j(\tau)\alpha_j(\tau)\right)}\,,
\label{intrealfermion}
\end{eqnarray}
and since each pair of Grassmann variables $\alpha_i\beta^*_i$ and
$\alpha_j\beta^*_j$ commutes, we have
$(\alpha_i\beta^*_i)(\alpha_j\beta^*_j)=(\alpha_j\beta^*_j)(\alpha_i\beta^*_i)$.
Therefore in the functional version of the model we can substitute
the Gaussian integral given in Eq. (\ref{intreal}) in Eq.
(\ref{intrealfermion}). We get
\begin{eqnarray}
R=a_0\,(\det{H})^{-1}\int [dr]
e^{\sum_{i,j=1}^N\int_0^{\beta}d\tau\,
r^*_i(\tau)H^{-1}_{ij}r_j(\tau)+\frac{1}{\sqrt{N}}\sum_{i=1}^N
\int_0^{\beta}d\tau\,\Bigl(r_i(\tau)\alpha_i^*(\tau)\beta_i(\tau)\,
+\,r^*_i(\tau)\beta^*_i(\tau)\alpha_i(\tau)\Bigr)}\,,
\label{intrealf1}
\end{eqnarray}
where the auxiliary fields $r_i(\tau)$ satisfy periodic boundary
conditions, i.e., $r_i(0)=r_i(\beta)$ and also
$r_i^*(0)=r_i^*(\beta)$. In the above expression the quantity
$a_{0}$ is a numerical factor that can be absorbed in the
normalization factor and the term
$[dr]\equiv\prod_i^N[dr_i][dr^*_i]$ is the functional measure for
the auxiliary complex fields.
Substituting the Eq. (\ref{intrealf1}) in the partition function
$Z$, we obtain
\begin{eqnarray}
Z=a_0\,(\det{H})^{-1}\int [dr]
e^{-\sum_{i,j=1}^N\int_0^{\beta}d\tau\,
r^*_i(\tau)H^{-1}_{ij}r_j(\tau)}\,\int [d\eta]\,e^{\,S_r}\,,
\label{part1}
\end{eqnarray}
where in Eq. (\ref{part1}) we have $S_r\equiv
S_r(r_i,r^*_i,b,b^*,\alpha_i,...)$.  Consequently the last
functional integral $\int [d\eta]\,e^{\,S_r}$ depends on the
auxiliary fields $r^*_i(\tau)$ and $r_i(\tau)$. Moreover, the
action $S_r$ can be separated into a free action for the bosons
and a Gaussian fermionic part. The action $S_r$ can be written in
the form

\begin{equation}
S_r = S_{0}(b,b^*) +  \int_{0}^{\beta} d\tau\,\sum_{i=1}^{N}\,
\rho^{\dagger}_{i}(\tau)\,M_r\,\rho_{i}(\tau)\,.
\label{pseudoacao}
\end{equation}
We would like to stress that in Eq. (\ref{pseudoacao}), instead of
writing $M(r_i,r^*_i,b^{*},b)$, for simplicity we use the notation
$M_r$, i.e., $M_r\equiv M(r_i,r^*_i,b^{*},b)$. The column matrix
$\rho_{\,i}(\tau)$ is given in terms of the Grassmann  Fermi
fields
\begin{eqnarray}
\rho_{\,i}(\tau) &=& \left(
\begin{array}{c}
\beta_{\,i}(\tau) \\
\alpha_{\,i}(\tau)
\end{array}
\right),
\nonumber\\
\rho^{\dagger}_{\,i}(\tau) &=& \left(
\begin{array}{cc}
\beta^{*}_{\,i}(\tau) & \alpha^{*}_{\,i}(\tau)
\end{array}
\right) \label{69a}
\end{eqnarray}
and the matrix $M_r$ is given by
\begin{equation}
M_r= \left( \begin{array}{cc}
\partial_{\tau} + \frac{\Omega}{2} &  -N^{-1/2}\,\biggl(g_{1}
\,b^{*}\,(\tau) + g_{2}\,b\,(\tau)-r^*_i(\tau)\biggr) \\
-N^{-1/2}\,\biggl(g_{1}\,b\,(\tau) + g_{2}\,b^{*}\,(\tau)-r_i(\tau)\biggr)
& \partial_{\tau} - \frac{\Omega}{2}
\end{array} \right)\, .
\label{matrix1}
\end{equation}
The complex functions and Grassmann Fermi fields $r_i(\tau)$,
$b(\tau)$, $\alpha_{i}(\tau)$ and $\beta_{i}(\tau)$ can be
represented in terms of a Fourier expansion. Therefore, we have
\begin{equation}
r_i(\tau) = \beta^{-1/2} \sum_{v} r_i(v)\, e^{i v
\tau}\,,
 \label{69r}
\end{equation}
\begin{equation}
b(\tau) = \beta^{-1/2} \sum_{\omega} b(\omega)\, e^{i\omega
\tau}\,,
 \label{69c}
\end{equation}
and
\begin{equation}
\rho_{i}(\tau) = \beta^{-1/2} \sum_{p} \rho_{i}(p)\, e^{ip \tau}\,
. \label{71}
\end{equation}
Since the complex functions $r_i(\tau)$ and $b(\tau)$ obey
periodic boundary conditions, and the Grassmann Fermi fields
$\alpha_{i}(\tau)$ and $\beta_{i}(\tau)$ obey anti-periodic
boundary conditions, we have that $v= \frac{2\pi n}{\beta}$,
$\omega = \frac{2\pi n}{\beta}$ and $p=\frac{(2n+1)\pi}{\beta}$,
for integer $n$. They are respectively the bosonic and fermionic
Matsubara frequencies.
Substituting the Fourier expansions in the action given by Eq.
(\ref{pseudoacao}) we get
\begin{equation}
S_r = \sum_{\omega} (i\omega -
\omega_{0})\,b^{*}(\omega)\,b(\omega) + \sum_{p,\,q}
\sum_{i=1}^{N}\, \rho^{\dagger}_{i}(p)\, M_{p\,q}\,\rho_{i}(q)\, ,
\label{72b}
\end{equation}
where the matrix $M_{p\,q}\equiv M_{p\,q}(r_i,r_i^*,b,b^*)$ is
given by
\begin{equation}
M_{p\,q} = \left( \begin{array}{cc}
(ip + \Omega/2)\delta_{p\,q} & -(N\,\beta)^{-1/2}\,Q^{\dagger}_{p\,q}\\
-(N\,\beta)^{-1/2}\,Q_{p\,q} &
(ip - \Omega/2)\delta_{p\,q}
\end{array} \right)
\label{matrix2}
\end{equation}
and the function $Q_{p\,q}\equiv Q_{p\,q}(r_i,b,b^*)$ can be
written as
\begin{equation}
Q_{p\,q} = g_{1}\,b(p-q) + g_{2}\,b^{*}(q-p)-r_i(p-q)\,.
\label{Qdef}
\end{equation}
Since the integrals with respect to the Fermi fields are Gaussian,
we may integrate over these Grassmann variables. This procedure
yields
\begin{equation}
\int[\,d\eta(\rho)]\exp\biggl(\sum_{p,q}
\sum_{i=1}^{N}\,\rho^{\dagger}_{i}(p)\,
M_{p\,q}\,\rho_{i}(q)\biggr)= \prod_{i=1}^{N}\,\det M_i\, ,
\label{intgrass}
\end{equation}
where the matrix $M_i\equiv M_i(r_i,r^*_i,b^{*},b)$ is a block
matrix of the following form
\begin{equation}
M_i = \left( \begin{array}{cc}
iP + \frac{\Omega}{2}\,I & -(N\beta)^{-1/2}\,Q^{\dagger}\\
-(N\beta)^{-1/2}\,Q & iP - \frac{\Omega}{2}\,I
\end{array} \right)\,.
\label{77b}
\end{equation}
In the above equation $I$ is the identity matrix and the
components of matrix $P$ are $P_{p\,q}=p\,\delta_{p\,q}$ and $Q$
was defined in Eq. (\ref{Qdef}). Our aim is to present the
quotient $\frac{Z}{Z_0}$, defined in Eq. (\ref{partition1}).
Combining the results obtained in Eq. (\ref{part1}), Eq.
(\ref{pseudoacao}) and Eq. (\ref{intgrass}) we get that
$\frac{Z}{Z_{0}}$ is given by
\begin{equation}
\frac{(\det{H})^{-1}\int [\,dr] e^{\sum_{i,j=1}^N\sum_v
r^*_i(v)H^{-1}_{ij}r_j(v)}\,\int [\,d\eta(b)]e^{\sum_{\omega}
(i\omega - \omega_{0})b^{*}(\omega)b(\omega)}\prod_{j=1}^N\det
M_j} {\int [\,d\eta(b)] e^{\sum_{\omega}(i\omega -
\omega_{0})b^{*}(\omega)b(\omega)} {\det}^NM(0,0)} \label{b74}
\end{equation}
where the functional measure $[\,d\eta(b)]$ is defined by
\begin{equation}
[\,d\eta(b)]=\prod_{\omega}db(\omega)\,db^{*}(\omega)\,.
\label{b75}
\end{equation}
In the Eq. (\ref{b74}) the matrix $M(0,0)$ is given by
\begin{equation}
M(0,0) = \left( \begin{array}{cc}
iP + \frac{\Omega}{2}\,I & 0\\
0 & iP - \frac{\Omega}{2}\,I
\end{array} \right)\,.
\label{79b}
\end{equation}
In order to simplify the calculations, let us change variables in
the following way:
\begin{equation}
b(\omega)\rightarrow~\Biggl(\frac{\pi}{(\omega_{0} -
i\omega)}\Biggr)^{1/2}b(\omega) \label{78}
\end{equation}
and
\begin{equation}
b^{*}(\omega)\rightarrow\Biggr(\frac{\pi}{(\omega_{0} -
i\omega)}\Biggr)^{1/2}b^{*}(\omega)\, . \label{79}
\end{equation}
Note that Eq. (\ref{79}) is not the complex conjugate of Eq.
(\ref{78}). It is not difficult to see that after these changes of
variables, the denominator of the Eq. (\ref{b74}) turns out to be
equal to unity
\begin{equation}
\int [\,d\eta(b)] \exp\biggl(-\pi
\sum_{\omega}b^{*}(\omega)b(\omega)\biggr) = 1\,. \label{80}
\end{equation}
We can express the ratio $\frac{Z}{Z_{0}}$ by the integral
\begin{equation}
\frac{Z}{Z_{0}}=(\det{H})^{-1}\int [\,dr] \,e^{\sum_{i,j=1}^N
\sum_v r^*_i(v)H^{-1}_{ij}r_j(v)}\,\int[\,d\eta(b)]\,
\exp{\biggl(\,S_{\,eff}\,(b)\,\biggr)}\,, \label{81}
\end{equation}
where after doing the Fourier transform for the auxiliary fields
the functional measure becomes
$[\,dr]=\prod_i^N\prod_{\omega}dr_i(\omega)dr^*_i(\omega)$. The
effective action of the bosonic mode, $S_{\,eff}(b)$ is given by
\begin{equation}
S_{\,eff}\,=\,-\,\pi\,\sum_{\omega}\,b^{\,*}(\omega)\,
b(\omega)\,+\,\sum_{i=1}^N\ln{det\,(I\,+\,A_i)}\,. \label{81a}
\end{equation}
The matrix $A_i$ in the determinant of the above equation is given
by
\begin{equation}
\det(I+A_i) =
\det\biggl(M^{-1/2}(0,0)M_i(b^{*},b)M^{-1/2}(0,0)\biggr).
\label{82a}
\end{equation}
Performing the product of matrices we obtain the matrix $A$
afterwards. Therefore we can write
\begin{equation}
A_i = \left( \begin{array}{cc}
0 & B_i \\
-C_i & 0
\end{array} \right)\,.
\end{equation}
The components of the matrices $B_i$ and $C_i$ are given by
\begin{eqnarray}
({B_i})_{p\, q}&=&-\,\Biggl(\frac{1}{\beta
N}\Biggr)^{\frac{1}{2}}\Biggl(iq-\frac{\Omega}{2}
\Biggr)^{-\frac{1}{2}}\Biggl(ip+\frac{\Omega}{2}\Biggr)^{-\frac{1}{2}}\times\nonumber\\
&\times&\Biggl(\,\frac{\sqrt{\pi}g_{\,1}\,b^{*}\,(q-p)}{\sqrt{\omega_{0}-i(q-p)}}
+\frac{\sqrt{\pi}g_{\,2}\,b\,(p-q)}{\sqrt{\omega_{0}-i(p-q)}}-r_i^*(q-p)\Biggr)
\label{83a}
\end{eqnarray}
and
\begin{eqnarray}
({C_i})_{p\, q}&=&\Biggl(\frac{1}{\beta
N}\Biggr)^{\frac{1}{2}}\Biggl(ip-\frac{\Omega}{2}
\Biggr)^{-\frac{1}{2}}\Biggl(iq+\frac{\Omega}{2}
\Biggr)^{-\frac{1}{2}}\times\nonumber\\
&\times&\Biggl(\,\frac{\sqrt{\pi}g_{\,1}\,b\,(p-q)}{\sqrt{\omega_{0}-i(p-q)}}
+\frac{\sqrt{\pi}g_{\,2}\,b^{*}\,(q-p)}{\sqrt
{\omega_{0}-i(q-p)}}-r_i(p-q)\Biggr)\, . \label{83b}
\end{eqnarray}
In order to perform the functional integral given by Eq.
(\ref{81}) we must to find a manageable expression for
$det\,(I\,+\,A)$. For this case we can use the following identity
\begin{equation}
\det\,(I+A_i) = \det\,(I+B_iC_i)\rightarrow
\exp\biggl(tr(B_iC_i)\biggr)\, . \label{detaprox}
\end{equation}
So using this approximation given by Eq. (\ref{detaprox}) we can
find an expression for the ratio $\frac{Z}{Z_0}$ defined in Eq.
(\ref{81}) of the following form
\begin{eqnarray}
\frac{Z}{Z_0}\,=\,(\det{H})^{-1}\int [\,dr]
e^{\sum_{i,j=1}^N\sum_{v}
r_i(v)\,\left(H^{-1}_{ij}-\frac{1}{N}\tanh
\left(\frac{\beta\,\Omega}{4}\right)\,\frac{\delta_{ij}}{iv-
\Omega}\right)\,r^*_j(v)}\,\,\frac{Z_r}{Z_0}\,, \label{partition5}
\end{eqnarray}
where
\begin{eqnarray}
\frac{Z_r}{Z_0}&=&\int\,[d\eta(b)]\,e^{-\,\pi\,\sum_{\omega}\,b^{\,*}
(\omega)\,\Bigl(1\,-\,a(\omega)\,\Bigr)\,b(\omega)
\,+\,\pi\,\sum_{\omega}\,\Bigl(b(\omega)\,c(\omega)\,b(\,-\,\omega)\,+\,
b^{\,*}(\omega)\,c(\omega)\,b^{\,*}(-\,\omega)\,\Bigr)}\,\times\nonumber\\
&&\times\, e^{\frac{\pi}{N}\,\sum_{\omega}\,
\Bigl(d_1(\omega)\,b(\omega)\,+\,d_2(\omega)\,b^*(\omega)\,\Bigr)}\,.
\label{89}
\end{eqnarray}
In the Eq. (\ref{partition5}) the coefficients $a(\omega)$ and
$c(\omega)$ are given respectively by
\begin{equation}
a(\omega)\,=\,\Biggl(\frac{g_{\,1}^{\,2}\,(\Omega
-i\omega)^{\,-1}+\,g_{\,2}^{\,2}\,(\Omega\,+\,
i\omega)^{\,-1}}{(\omega_{0}\,-\,i\,\omega)}
\Biggr)\,\tanh{\biggl(\,\frac{\beta\,\Omega}{4}\,\biggr)}\,,
\label{90}
\end{equation}
\begin{equation}
c(\omega)\,=\,\Biggl(\frac{g_{\,1}\,g_{\,2}\,\Omega}{(\omega_{\,0}^{\,2}\,+
\,\omega^{\,2})^{\,1/2}\,(\Omega^{\,2}\,+\,\omega^{\,2})}\Biggr)
\,\tanh{\,\biggl(\frac{\beta\,\Omega}{4}\biggr)}\,,
 \label{90a}
\end{equation}
\begin{equation}
d_1(\omega)\,=\,-\frac{1}{\sqrt{\pi}\sqrt{\omega_
{0}\,-\,i\,\omega}}\Biggl(\frac{g_1}{\Omega
-i\omega}\sum_ir^*_i(\omega)+\frac{g_2}{\Omega\,+\,
i\omega}\sum_ir_i(-\omega)\Biggr)\,
\tanh{\biggl(\,\frac{\beta\,\Omega}{4}\,\biggr)}
 \label{90b}
\end{equation}
and
\begin{equation}
d_2(\omega)\,=\,-\frac{1}{\sqrt{\pi}\sqrt{\omega_
{0}\,-\,i\,\omega}}\Biggl(\frac{g_1}{\Omega
-i\omega}\sum_ir_i(\omega)+\frac{g_2}{\Omega\,+\,
i\omega}\sum_ir^*_i(-\omega)\Biggr)\,
\tanh{\biggl(\,\frac{\beta\,\Omega}{4}\,\biggr)}\,.
 \label{90c}
\end{equation}
Performing the Gaussian integral in the bosonic variables in the
Eq. (\ref{89}), we finally obtain
\begin{eqnarray}
\frac{Z_r}{Z_0}&=&I_0\,\times\,\exp\left(\frac{1}{N^2}\sum_{\omega}
\,\frac{A_1(\omega)}{\Bigl(1-a(-\omega)\Bigr)\,\Bigl(1-a(\omega)\Bigr)-
4\,c^2(\omega)}\,\sum_{ij}r_i(\omega)
\,r^*_j(\omega)\right.+\nonumber\\
&+&\left.\sum_{\omega}\,\frac{A_2(\omega)}{\Bigl
(1-a(-\omega)\Bigr)\,\Bigl(1-a(\omega)\Bigr)-
4\,c^2(\omega)}\,\left(\sum_{ij}r_i
(\omega)\,r_j(-\omega)\,+\,\sum_{ij}r^*_i
(\omega)\,r^*_j(-\omega)\right)\,\right)\,, \label{100}
\end{eqnarray}
where $I_0$, $A_1(\omega)$ and $A_2(\omega)$ are defined
respectively by
\begin{eqnarray}
I_0&=&\frac{1}{\Bigl(1-a(0)-2\,c(0)\Bigr)^{1/2}
\Bigl(1-a(0)+2\,c(0)\Bigr)^{1/2}}\,\times\nonumber\\
&&\times\,\prod_{\omega}\frac{1}{\Bigl(1-a(-\omega)
\Bigr)\Bigl(1-a(\omega)\Bigr)-4\,c^2(\omega)}\,,\nonumber\\
A_1(\omega)&=&\frac{\tanh{\biggl(\frac{\beta\,\Omega}{4}\biggr)}}{\pi
\Bigl(\Omega-i\omega\Bigr)^2}
\left(\frac{g_1^2}{\omega_0-i\omega}+\frac{g_2^2}{\omega_0+i\omega}-
\tanh{\biggl(\frac{\beta\,\Omega}{4}\biggr)}\frac{(g_1^2-g_2^2)^2}{(\omega_0^2-
\omega^2)(\Omega+i\omega)}\right)\,,\nonumber\\
A_2(\omega)&=&\frac{g_1\,g_2\, \omega_0}{\pi\,(\omega_0^2-
\omega^2)(\Omega^2+\omega^2)}\tanh{\biggl(\frac{\beta\,\Omega}{4}\biggr)}\,.
\label{101}
\end{eqnarray}
Comparing the partition function for the full Dicke model in Eq.
(\ref{95}), with the partition function of the model with
dipole-dipole coupling in Eqs. (\ref{partition5}) and (\ref{100}),
we can see that the poles in the partition function are the same
and therefore the temperature of the phase transition from normal
to super-radiant state and the presence of a condensate is not
affected by the dipole-dipole coupling. The energy level spectrum
of the collective bosonic excitations of the model is the same as
the original full Dicke model. The quantum phase transition is not
modified in this situation. An important question here, is why the
full Dicke model with the dipole-dipole interaction has the same
critical behavior as the original model?

\section{Conclusions}\

With the development of quantum information and its application in
computation and communication the entangled states have been
attracted enormous interest, since several quantum protocols can
be realized exclusively with the help of entangled states. For
example new features in cryptography becomes possible utilizing
the non-classical properties of entangled pairs of particles. The
basic problem that arises in this area of research is how to
measuring entanglement in many-body systems. The Schmidt
decomposition can be used to measure correlation between two
sub-systems in a joint pure state. An important property is that
the reduced matrices (see appendix A) of both sub-systems written
in the Schmidt basis are diagonal and have the same positive
spectrum. For more than two entangled sub-systems the Schmidt
decomposition is in general impossible to be implemented. For
mixed states of two or more subsystems there is not a Schmidt
decomposition, and different measures of entanglement are
nonequivalent. The structure of entanglement in many-body systems
is much more difficult to study then for bipartite systems.

In a system of $N$ atoms, the presence of the dipole-dipole
interaction generates entangled states. If entangled has been
created in a portion of the many-body system, the critical
behavior of the system can changes revealing the strong
correlation between the atoms. In this work we investigate if the
entanglement between the atoms, generated by the dipole-dipole
interaction,  changes the critical properties of the system, as
the critical temperature that characterize the phase transition
from fluorescent to super-radiant phase and also the spectrum of
the bosonic excitation of the model.

First we consider the full Dicke model composed by a single
bosonic mode and an ensemble of $N$ identical two-level atoms.
Assuming that the system is in thermal equilibrium with a
reservoir at temperature $\beta^{-1}$, we consider the situation
where the coupling between the bosonic mode and the atoms
generates resonant and non-resonant processes. Secondly we
consider the full Dicke spin-boson model with the dipole-dipole
interaction between the atoms. We show that, introducing the
dipole-dipole interaction term, the critical temperature that
characterize the phase transition from fluorescent to
super-radiant phase does not change. Also, the spectrum of the
bosonic excitations of the model is unaffected by the
dipole-dipole interaction. Finally, the quantum phase transition
is not modified by the dipole-dipole interaction. Furthermore, as
discussed previously there is a zero temperature phase transition
when $g_{1}=0$ and $g_{2}\neq 0$. In this case there are no
thermal excitations and the phase transition is driven by the
quantum fluctuations. An important point is that there is a close
connection between quantum phase transition and entanglement. A
quantum phase transition can be characterized by long range
correlations. As the same way, the definition of entangled pure
state is that in the system at least one pair of the observables
has quantum correlations.

We might ask why we have obtained this result, since we expect at
principle that the non-Gaussian terms in the Hamiltonian can
modify phase transition properties. Remember that the spin-boson
Hamiltonian is a particular case of the Hamiltonian describing a
system of bound charges and a radiation field in the dipole
approximation \cite{power}, where we are assuming that the system
is confined in a region whose linear dimensions are small compared
with the wavelength of the bosonic mode. In the full Dicke model,
field non-uniformity is disregarded. All atoms are in the same
environment. The introduction of the dipole-dipole interaction
with infinite range does not modify this situation. Each two-level
atoms is coupled to infinitely many others. The above discussion
leads us to conclude that we are solving mean-field models. Other
heuristic argument is related to the fact that mean field theories
give exactly the right description of critical properties for
large $d$, where $d$ is the dimensionality of space. Since our
results are independent of the dimensionality, we are solving mean
field models. That is the reason why the dipole-dipole interaction
is not able to modify the critical properties for this cooperative
phenomena of super-radiance.

We conjecture that the critical properties are different if we go
beyond the mean field approximation. First let us assume that
linear dimensions of the system are not small compared with the
wavelength associated to the bosonic mode. In this case we have to
take into account the spatial variation in the coupling
coefficient between the pseudo-spin operators $\sigma_{(j)}^{+}$,
$\sigma_{(j)}^{-}$ and the boson annihilation and creation
operators of mode excitations. We note that Li and co-workers
\cite{li} investigated the quantum phase transition and the
super-radiant phase in the full Dicke model assuming that the
dimensions of the atomic ensemble are much larger than the
wavelength of the bosonic mode. This system has a critical
temperature of the phase transition from normal to super-radiant
state and also the presence of a condensate. With the introduction
of the dipole-dipole term, the critical temperature of the
super-radiant phase transition must change. This subject is under
investigation by the authors.

\section{Acknowlegements}

N. F. Svaiter would like to acknowledge the hospitality of the
Departamento de F\'{\i}sica, Universidade Federal da Para\'{\i}ba,
where part of this paper was carried out. This paper was supported
in part by Conselho Nacional de Desenvolvimento Cientifico e
Tecnol{\'o}gico do Brazil (CNPq).

\begin{appendix}
\makeatletter \@addtoreset{equation}{section} \makeatother
\renewcommand{\theequation}{\thesection.\arabic{equation}}

\section{The theory of pure and mixed states and the
reduced density operator.}\

In the theory of quantum systems, the notion of composite quantum
systems is fundamental. The Hilbert space of a composite quantum
system is the tensor product space of the Hilbert spaces
describing its sub-systems. Being more specific, let us consider a
bi-partite quantum system $S$ made of two quantum systems
$S^{(1)}$ and $S^{(2)}$, with respective Hilbert spaces ${\cal
H}^{(1)}$ and ${\cal H}^{(2)}$. The state space ${\cal H}$ of the
combined system $(S^{(1)}+S^{(2)})$ is given by the tensor product
of the Hilbert spaces pertaining to the sub-systems $S^{(1)}$ and
$S^{(2)}$. Therefore we have that the total Hilbert space can be
written as
\begin{equation}
{\cal H}={\cal H}^{(1)}\,\otimes\,{\cal H}^{(2)}. \label{1}
\end{equation}
If we take fixed orthonormal basis
$\{|\,\varphi_{_{i}}\,^{(1)}\,\rangle\}$ and
$\{|\,\varphi_{_{j}}\,^{(2)}\,\rangle\}$ in ${\cal H}^{(1)}$ and
${\cal H}^{(2)}$ respectively, a general state $|\psi\,\rangle$ in
the tensor product space ${\cal H}$ may be written as
\begin{equation}
|\,\psi\,\rangle=\sum_{i,\,j}\,\alpha\,_{i\,j}\,
|\,\varphi_{_{i}}\,^{(1)}\,\rangle\,\otimes
|\,\varphi_{_{j}}\,^{(2)}\,\rangle. \label{2}
\end{equation}
where $\{|\,\varphi_{_{i}}\,^{(1)}\,\rangle\,\otimes
|\,\varphi_{_{j}}\,^{(2)}\,\rangle\}$ is a basis of the tensor
product space.

Suppose that we have an operator $A\,^{(1)}$ acting only in ${\cal
H}^{(1)}$ and an operator $A\,^{(2)}$ acting only in ${\cal
H}^{(2)}$. Therefore we define the tensor product
$A\,^{(1)}\,\otimes\,A\,^{(2)}$ showing how does this quantity
acts on any element of the basis of the tensor product space. We
have
\begin{equation}
\left(A\,^{(1)}\,\otimes\,A\,^{(2)}\right)
\left(|\,\varphi_{_{i}}\,^{(1)}\,\rangle\,\otimes
|\,\varphi_{_{j}}\,^{(2)}\,\rangle\right)=
\left(A\,^{(1)}\,|\,\varphi_{_{i}}\,^{(1)}\,\rangle\right)\,
\otimes\,\left(A\,^{(2)}\,|\,\varphi_{_{j}}\,^{(2)}\,\rangle\right).
\label{3}
\end{equation}
This important definition lead us to its linear extension for an
arbitrary state  $|\psi\,\rangle$ . Using the Eq. (\ref{2}) and
the Eq. (\ref{3}) we get
\begin{equation}
\left(A\,^{(1)}\,\otimes\,A\,^{(2)}\right)|\psi\,\rangle=
 \sum_{i,\,j}\,\alpha\,_{i\,j}
\left(A\,^{(1)}|\,\varphi_{_{i}}\,^{(1)}\,\rangle\,\right)\,\otimes\,
\left(A\,^{(2)}\,|\,\varphi_{_{j}}\,^{(2)}\,\rangle\right).
\label{4}
\end{equation}
We also have that any product of operators acting in ${\cal H}$
can be represented as a linear combination of tensor products. We
have
\begin{equation}
A=\sum_{\alpha}\,\left(\,A\,_{\alpha}\,^{(1)}\,\otimes\,
A\,_{\alpha}\,^{(2)}\,\right). \label{5}
\end{equation}
The observable of the sub-system $S\,^{(1)}$ take the form
$A\,^{(1)}\,\otimes\,I\,^{(2)}$, while observable of sub-system
$S\,^{(2)}$ take the form $I\,^{(1)}\,\otimes\,A\,^{(2)}$.

Suppose that we are interested in the behavior of the sub-system
$S^{(1)}$. This behavior is determined by the expectation values
of the operators $\{\,X_{_{\alpha}}^{(1)}\,\}$, where the label
$(1)$ means that each operator acts on the states of the subsystem
$S^{(1)}$ alone, and the index $\alpha=1,2,...,N$ means operators
in ${\cal H}^{(1)}$. In other words, the label $\alpha$ denotes
the family of commutating operators
$\left(\,[X_{\alpha}\,^{(1)},X_{\beta}\,^{(1)}]=0\right)$, acting
in the state $|\,\psi\,\rangle$. Since the state vector
 $|\,\psi\,\rangle$ of the combined system can be expanded
using the orthonormal basis
$\{|\,\varphi_{_{i}}\,^{(1)}\,\rangle\}$ and
$\{|\,\varphi_{_{j}}\,^{(2)}\,\rangle\}$ we get for any operator
$\{\,X^{(1)}\,\}$, is its expectation value given by
\begin{equation}
\langle\,X^{(1)}\rangle=\langle\,\,\psi|\,X^{(1)}\,|
\psi\,\rangle. \label{8}
\end{equation}
Note that we drop the indices $\alpha$ in the operator. Using the
Eq. (\ref{2}), the fact that  the operator $X^{(1)}$ does not act
on the vectors that span the Hilbert space ${\cal H}^{(2)}$, and
the orthonormality of the
$\{\,|\,\varphi_{_{j}}\,^{(2)}\,\rangle\,\}$ basis, we get
\begin{equation}
\,\langle\,X^{(1)}\,\rangle=\sum_{ijk}\,\alpha_{kj}^{*}\alpha_{ij}
\langle\,\varphi_{_{k}}\,^{(1)}\,|\,X^{(1)}\,|\,\varphi_{_{i}}\,^{(1)}\rangle.
\label{9}
\end{equation}
Defining $c_{ik}=\sum_{j}\,\alpha_{ij}\,\alpha_{kj}^{*}$, we have
that the expectation value of the operator $X^{(1)}$ in a general
state $| \psi\,\rangle$, defined by $\langle\,X^{(1)}\,\rangle$,
is given by
\begin{equation}
\langle\,X^{(1)}\,\rangle=\sum_{ik}\,c_{ik}
\langle\,\varphi_{_{k}}\,^{(1)}\,|X^{(1)}\,|\,\varphi_{_{i}}\,^{(1)}\rangle,
\label{10}
\end{equation}
or
\begin{equation}
\langle\,X^{(1)}\,\rangle= tr\,[X^{(1)}\rho^{(1)}\,], \label{11}
\end{equation}
where
\begin{equation}
\rho^{(1)}=\sum_{ik}\,c_{ik}|\,\varphi_{_{i}}\,^{(1)}\rangle\,\langle
\varphi_{_{k}}\,^{(1)}\,|. \label{12}
\end{equation}
The quantity $\rho\,^{(1)}$ is called the density operator of the
sub-system $S^{(1)}$. The density operator describes the state of
the system interacting with other system in the same way as the
state vector describes the state of an isolated system.

Is a fundamental interest to measuring entangled in many body
systems, where we have usually mixed sates.  For simplicity, let
us consider a two-component quantum system $S^{(1)}$ and $S^{(2)}$
in a pure state $|\,\Phi\,\rangle$. A pure state is separable if
and only if the reduced density operators represent pure states.
Essentially separable states satisfy the classical separability
principle. In a bipartite quantum system with an entangled state
$|\,\Phi_{s}\,\rangle$, measurements in its sub-systems show that
$|\,\Phi_{s}\,\rangle$ contain information in the measurements in
$S^{(1)}$ and $S^{(2)}$ and also correlation between the
measurements. To test whether the sub-systems $S^{(1)}$ and
$S^{(2)}$ are entangled or not, we perform the Schmidt
decomposition of the state vector $|\,\Phi\,\rangle$. If the
Schmidt number is greater then one, the state $|\,\Phi\,\rangle$
is an entangled state. Therefore a pure state is entangled if and
only of the reduced density operators for the sub-systems describe
mixed states. Using the Schmidt decomposition it is possible to
evaluate the density operators $\rho^{(1)}_{mixed}=Tr_{(2)}
|\,\Phi\,\rangle\,\langle \,\Phi\,|$ and
$\rho^{(2)}_{mixed}=Tr_{(1)} |\,\Phi\,\rangle\,\langle \,\Phi\,|$.
The two reduced density operators have the same eigenvalues, and
the two sub-systems have identical von Neumann entropies. On the
other hand, if the Schmidt number is one, the reduced density
operators for $S^{(1)}$ or $S^{(2)}$ represent pure states and the
bipartite system is not an entangled state. The von Neumann
entropy of a pure state is zero. Suppose the operators $X^{(1)}$
and $X^{(2)}$ defined as $X\,^{(1)}\,\otimes\,I\,^{(2)}$ and
$I\,^{(1)}\,\otimes\,X\,^{(2)}$. The fluctuation operator are
defined as $\delta\,X^{(1)}=X^{(1)}-\langle\,X^{(1)}\rangle$ and
$\delta\,X^{(2)}=X^{(2)}-\langle\,X^{(2)}\rangle$. In practice, a
pure state is separable if and only if the quantum fluctuations of
all observable (linear self-adjoint operators acting in the
respective Hilbert space of each sub-system) are uncorrelated. If
at least one pair of linear self-adjoint operators have correlated
quantum fluctuations, the pure state is entangled.

Note that the von Neumann entropy attains the maximum $S(\rho)= ln
\,N$, in the case of completely mixed state where
$\rho=\frac{1}{N}\sum_{_{i}}|\psi_{i} \rangle\,\langle\,
\psi_{i}|$, where $N$ is the dimension of the corresponding
Hilbert space. Thus quantity is a monotonically function of the
degree of entangled between a pair of sub-systems. Therefore the
von Neumannn entropy of the density matrix for $S^{(1)}$ or
$S^{(2)}$ of the bipartite system is a measurement of entangled
\cite{amico}.

\end{appendix}


\begin{thebibliography}{99}

\bibitem{separability} {\em{''Quantum Theory and Pictures of Reality"}},
W. Schommers (Editor), Springer Verlag, Berlin Heidelberg (1989).
\bibitem{separability2} B. d'Espagnat, Phys. Rep. {\bf 110}, 202
(1984).
\bibitem{schrodinger} E. Schr\"odinger,
Naturwissenschaften {\bf 23}, 807 (1935).
\bibitem{light0} J. F. Clauser, M. A. Horne, A. Shimony and R. A. Holt,
Phys. Rev. Lett. {\bf 23}, 880 (1969).
\bibitem{light} V. Vedral, M. A. Rippin and M. B. Plenio, J. Mod.
Opt. {\bf 44}, 2185 (1997).
\bibitem{bell1} J. S. Bell, Physics {\bf 1}, 195 (1965).
\bibitem{bell2} J. S. Bell, Rev. Mod. Phys. {\bf 38}, 447 (1966).
\bibitem{bell} J. S. Bell, {\em{"The Foundations of Quantum
Mechanics"}}, J. Bell, K. Gottfried and M. Veltman (Editors),
World Scientific (2001).
\bibitem{exp1} B. W. Schumacher, Phys. Rev. {\bf A44}, 7047
(1991).
\bibitem{exp2} N. Gisin, Phys. Lett. {\bf 154A}, 201 (1991).
\bibitem{exp3} R. R. Puri, J. Phys. {\bf A29}, 5719 (1996).
\bibitem{mann} A. Mann, M. Revzen and W. Schleich, Phys. Rev. {\bf
A46}, 5363 (1992).
\bibitem{zurek1} {\em{''Quantum Theory and Measurement"}}, J. A.
Wheeler and W. H. Zurek (Editors), Princeton University Press
(1983).
\bibitem{khalili} V. B. Braginsky and F. Y. Khalili,
{\em{"Quantum Measurement"}}, Cambridge University Press,
cambridge (1992).
\bibitem{namiki} M. Namiki, S. Pascazio and H.
Nakazato, {\em{"Decoherence and Quantum Measurements"}}, World
Scientific, Singapure (1997).
\bibitem{benioff} P. Benioff, Phys. Rev. Lett.
{\bf 48}, 1581 (1982).
\bibitem{fey1} R. P. Feynman, Int. Jour. Mod.
Phys. {\bf 21}, 467 (1982).
\bibitem{albert} D. Z. Albert, Phys. Lett.
{\bf A98}, 249 (1983).
\bibitem{deutsch} D. Deutsch, Proc. R. Soc. Lond. {\bf A400}, 97 (1985).
\bibitem{fey2} R. P. Feynman, Found. Phys. {\bf 16}, 507 (1986).
\bibitem{di} D. P. DiVincenzo, Phys. Rev. {\bf A51}, 1015 (1995).
\bibitem{steane} A. Steane, Rep. Prog. Theor. Phys. {\bf 61}, 117
(1998).
\bibitem{livro} {\em{''The Physics of Quantum Information"}}, D.
Bouwmeester, A. Ekert and A. Zeilinger (Editors), Springer Verlag,
Berlin (2000).
\bibitem{176} D. M. Greenberg, M. A. Horne, A.
Shimony and  A. Zeilinger, Am. J. Phys. {\bf 58}, 1131 (1990).
\bibitem{zurek} W. H. Zurek, Physics Today {\bf 44}, 36 (1991).
\bibitem{hep} F. Benatti and R. Floreani, {\em{"Open Quantum Dynamics: Complete
Positivity and Entanglement"}}, quant. ph/0507271.
\bibitem{rbrandes} T. Brandes, Phys. Rep. {\bf 408}, 315 (2005).
\bibitem{wiesner} C. H. Bennett and S. J. Wiesner, Phys. Rev. Lett.
{\bf 69}, 2881 (1992).
\bibitem{wootters} C. H. Bennett, G. Brassard, C. Crepeau, R.
Jozsa, A. Peres and W. K. Wootters, Phys. Rev. Lett. {\bf 70},
1895 (1993).
\bibitem{cirac} J. I. Cirac and P. Zoller, Phys. Rev. Lett.
{\bf 74}, 4091 (1995).
\bibitem{loss} D. Loss and D. P. DiVincenzo, Phys. Rev. {\bf A57},
120 (1998).
\bibitem{cm1} X. Wang, H. Fu and A. I. Solomon, J. Phys. {\bf
A34}, 11307 (2001).
\bibitem{cm2} M. C. Arnesen, S. Bose and V. Vedral, Phys. Rev.
Lett. {\bf 87}, 017901 (2001).
\bibitem{cm3} D. Gunlycke, V. M. Kendon, V. Vedral and S. Bose,
Phys. Rev. {\bf A64}, 042302 (2001).
\bibitem{lehmberg1} R. H.
Lehmberg, Phys. Rev. {\bf A2}, 883 (1970).
\bibitem{lehmberg2} R. H. Lehmberg, Phys. Rev. {\bf A2}, 889
(1970).
\bibitem{agarval} G. S. Agarval, A. C. Brown, L. M. Narducci and
G. Vetri, Phys. Rev. {\bf A15}, 1613 (1977).
\bibitem{tanas} Z. Ficek, R Tan\'as and S. Kielich, Physica {\bf A146}, 452 (1987).
\bibitem{ficek} T. G. Rudolph, Z. Ficek and B. J. Dalton,
Phys. Rev. {\bf A52}, 636 (1995).
\bibitem{ujihara} H. T. Dung and K. Ujihara, Phys. Rev. Lett. {\bf 84}, 254 (2000).
\bibitem{guo} G. C. Guo and c. P. Yang, Physica {\bf A260}, 173 (1998).
\bibitem{amico} L. Amico, R. Fazio, A. Osterloch and V. Vedral,
{\em{"Entanglement in Many-Body Systems"}}, quant-phys/0703044
(2007).
\bibitem{eb} C. Emary and T. Brandes, Phys. Rev. Lett. {\bf
90}, 044101-1 (2003).
\bibitem{eb2} C. Emary and T. Brandes, Phys. Rev. {\bf E67}, 066203 (2003).
\bibitem{brody} T. A. Brody, J. Flores, J. B. french, P. A. Melo,
A. Pandey and S. S. M. Wong, Rev. Mod. Phys. {\bf 53}, 385 (1981).
\bibitem{mehta} M. L. Mehta, {\em{''Random Matrices"}}, Elsevier,
Amsterdam (2004).
\bibitem{lambert2} N. Lambert, C. Emary and T. Brandes, Phys. Rev. {\bf A71}, 053804-1 (2005).
\bibitem{gross} M. Gross and S. Haroche, Phys. Rep. {\bf 93}, 301
(1982).
\bibitem{hertz} J. A. Hertz, Phys. Rev. {\bf B14}, 1165
(1976).
\bibitem{sachdev} S. Sachdev, {\em{''Quantum Phase Transitions"}},
Cambridge University Press, Cambridge (1999).
\bibitem{brennen} G. K. Brennen, I. H. Deutsch and P. S. Jessen, Phys. Rev. {\bf A61},
062309 (2000).
\bibitem{hartmann} F. Friedberg and S. R. Hartmann, Phys. Rev.
{\bf A10}, 1728 (1974).
\bibitem{coffey} B. Coffey and R. Friedberg, Phys. Rev. {\bf A17},
1033 (1978).
\bibitem{osborn} T. J. Osborne and M. A. Nielsen, Phys. Rev. {\bf A66},
032110 (2002).
\bibitem{amico1} A. Ostertoh, L. Amico, F. Plastina and R. Fazio,
Nature {\bf 416}, 608 (2002).
\bibitem{kitaev} G. Vidal, J. I. Latorre, E. Rico and A. Kitaev,
Phys. Rev. Lett. {\bf 90}, 227902 (2003).
\bibitem{dicke} R. H. Dicke, Phys.
Rev. {\bf 93}, 99 (1954).
\bibitem{hl1} K. Hepp and E. H. Lieb, Ann. Phys. {\bf 76}, 360
(1973).
\bibitem{wanghioe} Y. K. Wang and F. T. Hioe, Phys. Rev. {\bf A7},
831 (1973).
\bibitem{hl2} K. Hepp and E. H. Lieb, Phys. Rev. {\bf A8},
2517 (1973).
\bibitem{hioe} F. T. Hioe, Phys. Rev. {\bf A8}, 1440 (1973).
\bibitem{duncan} G. Comer Duncan, Rev. {\bf A8}, 418 (1974).
\bibitem{tese} M. Aparicio Alcalde, A. L. L. de
Lemos and N. F. Svaiter, J. Phys. {\bf A40}, 11961 (2007).
\bibitem{pizi} B. M. Pimentel and A. H. Zimerman, Nuovo Cim. {\bf
B30}, 43 (1975).
\bibitem{pizi2} B. M. Pimentel and A. H. Zimerman, Phys. Lett. {\bf A53},
200 (1975).
\bibitem{jc} E. T. Jaynes and F. W. Cummings, Proc. IEEE
{\bf 51}, 89 (1963).
\bibitem{artigo} M. Aparicio Alcalde, R. Kullock and N. F. Svaiter,
J. Math. Phys. {\bf 50}, 013511-1 (2009).
\bibitem{chang} L. D. Chang and S. Chakravarty, Phys. Rev.
{\bf B31}, 154 (1985).
\bibitem{legget} A. J. Legget, C. Chakravarty,
A. T. Dorsey, M. P. A. Fisher, A Garg and W. Zwerger, Rev. Mod.
Phys. {\bf 59}, 1 (1987).
\bibitem{parma} G. M. Parma, K,-A. Seiominen and A. K. Ekert,
Proc. R. Soc. Lond. {\bf A452}, 567 (1996).
\bibitem{benatti} J. H. Reina, L. Quiroga and N. F. Johnson, Phys. Rev. {\bf A65}, 032326
(2002).
\bibitem{jj1} B. Buck and C. V. Sukumar, Phys. Lett. {\bf A81},
132 (1981).
\bibitem{jj2} B. Buck and C. V. Sukumar, J. Phys. {\bf A17},
885 (1984).
\bibitem{jj3} V. Buzek, Phys. Rev. {\bf A39}, 3196 (1989).
\bibitem{puri} R. R. Puri, {\em{"Mathematical Methods of
Quantum Optics"}}, Springer-Verlag, Berlin Heidelberg (2001).
\bibitem{ja2} P. L. Knight and L. Allen, {\em{''Concepts in Quantum Optics"}},
 Pergamon Press Inc., New York (1983).
\bibitem{haldane} F. D. M. Haldane, Phys. Rev. Lett. {\bf 60}, 635
(1988).
\bibitem{pittel} J. Dukelky, S. Pittel and G. Sierra, Rev. Mod.
Phys. {\bf 76}, 643 (2004).
\bibitem{popov1} V. N. Popov and S. A. Fedotov, Theor. Math. Phys.
{\bf 51}, 363 (1982).
\bibitem{popov2} V. N. Popov and S. A. Fedotov, Sov. Phys. JETP {\bf 67}, 535 (1988).
\bibitem{power} E. A. Power and S. Zineau, Phyl. Trans. Roy. Soc.
{\bf 251}, 427 (1951).
\bibitem{li} Y. Li, Z. D. Wang and C. P. Sun, Phys. Rev. {\bf A74},
023815 (2006).


\end{thebibliography}
\end{document}